\documentclass[preprint2]{aastex}

\usepackage{lscape}
\usepackage{apjfonts}

\newcommand{\footstar}[1]{$^*$ \footnotetext{$^*$#1}}

\begin{document}


\title{Final Binary Stars Results From The VLT Lunar Occultations Program
\footstar{Based on observations made with ESO telescopes at Paranal Observatory}
}


\author{A. Richichi\altaffilmark{1}, O. Fors\altaffilmark{2,3,4}, 
F. Cusano\altaffilmark{5}, and V. Ivanov\altaffilmark{6}}


\altaffiltext{1}{National Astronomical Research Institute of Thailand, 191 Siriphanich Bldg., 
Huay Kaew Rd., Suthep, Muang, Chiang Mai  50200, Thailand}
\email{andrea4work@gmail.com}
\altaffiltext{2}{Departament Astronomia i Meteorologia and Institut de Ci\`encies del Cosmos (ICC), 
Universitat de Barcelona (UB/IEEC), Mart\'{\i} i Franqu\'es 1, 08028 Barcelona, Spain}
\altaffiltext{3}{Observatori Fabra, Cam\'{\i} de l'Observatori s/n, 08035, Barcelona, Spain}
\altaffiltext{4}{Qatar Environment and Energy Research Institute, Qatar Foundation, Tornado Tower, Floor 19, P.O. Box 5825, Doha, Qatar}
\altaffiltext{5}{INAF-Osservatorio Astronomico di Bologna, Via Ranzani 1, 40127 Bologna, Italy}
\altaffiltext{6}{European Southern Observatory, Ave. Alonso de Cordova 3107,
Casilla 19001, Santiago 19, Chile}


\begin{abstract}
We report on 13 sub-arcsecond binaries, detected by means of lunar occultations
in the near-infrared at the ESO Very Large Telescope.
They are all first-time detections, except for the visual binary
HD~158122 which we resolved for the first time in the near infrared.
The primaries have magnitudes in the range
$K$=4.5 to 10.0, and the companions in the range
$K$=6.8 to 11.1. The magnitude differences have a median
value of 2.4, with the largest being 4.6.
The projected separations are in the range 4 to 168 milliarcseconds 
with a median of 13 milliarcseconds. 
We discuss and compare our results with the available literature.

With this paper, we conclude the mining for binary star detections
in the volume of 1226 occultations recorded at the VLT with the ISAAC instrument. 
We expect that the majority of these binaries
may be unresolvable by adaptive optics on current telescopes,
and they might be challenging for long-baseline interferometry.
However they  constitute an interesting sample for future
larger telescopes and for astrometric missions such as GAIA.
\end{abstract}





\section{Introduction}
\label{section:introduction}
We continue our series of results from a routine program of
lunar occultations (LOs) observed in the near-infrared
with the ESO Very Large Telescope (VLT). The observational
data, the analysis and the type of results -- namely
binaries with small projected separations and largely
consisting in first-time detections --  follow closely those presented
in \citet[R13 hereafter]{2013AJ....146...59R} and
other papers referenced therein. Accordingly, we keep
the introduction and the data analysis to a minimum.

In Sect.~\ref{section:data} we briefly summarize the
observations and the data analysis.
In Sect.~\ref{section:results} we comment on the
list of sources and we discuss
the individual results, when possible in the context
of available bibliography. 

With this batch of results we have now
exhausted the LO observations
the VLT using the ISAAC instrument, for what concerns the detection
of binary stars. In Sect.~\ref{sec:conclusions} we
provide some statistical conclusions.

\section{Observations and data analysis}
\label{section:data}

The observations were carried out in 2012 from
April through October, and cover the
ESO Periods 89 and 90. We used the
8.2-m UT3 Melipal telescope of
the VLT and the ISAAC instrument 
\citep{1999Msngr..95....1M}
operated in burst mode.
In each period, 15 hours of
telescope time were allocated, and were used to
observe 161 and 130 occultation events, respectively, including
 pointing, acquisition and other overheads.
In fact, the P90 time allocation
started already before the end of P89, and the observations
were concluded early in the period given the pending
decommissioning of ISAAC. 

All observations were carried out in service mode,
based on a long list of potentially observable events
from which the actual targets were later chosen
by the Observatory staff depending on the availability
of suitable time slots. As a result,
the sources were 
inherently selected at random. As expected, the majority
were found to be unresolved: in a previous paper
\citep{apjssur} we listed those observed until June 2012.
A few sources were found to be resolved or extended,
and will be the subject of future publications.
Here we deal with those found to be binaries.
They are listed
in Table~\ref{tab:list}, with the same format used in
R13. They are denoted by a sequential number for ease
of reference across this paper, and by a 
field "Our ID" which includes the ESO Period and
the designation in our database.

Our predictions were generated from
the 2MASS Catalogue, from which also the near-infrared
photometry listed in Table~\ref{tab:list} is 
drawn. We did not attempt to derive 
proper $K$-band photometry from our light curves, but
we compared our 
measured counts with those predicted
based on the 2MASS magnitude. Cases in which
a significant discrepancy was present are
noted in Sect.~\ref{section:results}.

Each observation consisted of 7500 frames
in a 32x32-pixel ($4\farcs7 \times 4\farcs7$) sub-window,
with a time sampling of 3.2\,ms. This was also the effective
integration time. A broad-band $K_{\rm s}$ filter was
employed for all events, except in the case of
the bright Star~6,
for which
a narrow-band filter centered at 2.07\,$\mu$m was
employed to avoid possible non-linearities. 
The events were disappearances, with lunar
phases ranging from 35\% to 97\% (median 68\%). 
Airmass ranged  
from 1.0 to 2.0, while seeing ranged
from  $0\farcs6$ to $1\farcs2$ (median $0\farcs8$). 
In any  case the LO light curves are largely insensitive to atmospheric
perturbations.

Details on the conversion of the data cubes to light curves
and on the data analysis
have been given in R13 and references therein.
It should be noted that in general we restrict our analysis
to $\approx 0\farcs5$ around the main event. Therefore,
any companions with  projected separations larger than this would not appear
among our results.

\begin{deluxetable}{rllllrrrrrrl}
\rotate
\tabletypesize{\small}
\tablecaption{List of detected binaries\label{tab:list}}
\tablewidth{0pt}
\tablehead{
\colhead{Seq}&
\colhead{Our ID}&
\colhead{2MASS ID}&
\colhead{Simbad ID}&
\colhead{Date}&\colhead{UT}&
\colhead{$B$}&\colhead{$V$}&
\colhead{$J$}&\colhead{$H$}&\colhead{$K$}&
\colhead{Sp.}
}
\startdata
1	&	P89-001	&	10304197+0354115	&	BD+04 2345	&	2012 April 04	&	04:15:03	&	10.19	&	9.75	&	8.77	&	8.58	&	8.49	&	 F8 	\\
2	&	P89-040	&	17121145-2134332	&	IRAS 17091-2130	&	2012 July 02	&	05:29:17	&	  	&		&	7.08	&	6.03	&	5.60	&		\\
3	&	P89-042	&	17123952-2136216	&	HD 155469	&	2012 July 02	&	05:45:02	&	7.40	&	7.03	&	6.23	&	6.07	&	6.00	&	 F2V 	\\
4	&	P89-055	&	12460117-0918482	&	LTT 4882	&	2012 July 25	&	00:47:16	&	8.74	&	8.05	&	6.88	&	6.58	&	6.51	&	 G4 	\\
5	&	P89-056	&	12480144-0908398	&	HD 111241	&	2012 July 25	&	01:22:37	&	9.23	&	8.85	&	8.04	&	7.84	&	7.76	&	 F0 	\\
6	&	P89-102	&	17262278-2122494	&	TYC 6246-232-1	&	2012 August 26	&	00:55:45	&	12.18	&	10.16	&	5.76	&	4.82	&	4.46	&		\\
7	&	P89-114	&	17281619-2057493	&	HD 158122	&	2012 August 26	&	02:17:30	&	8.37	&	7.94	&	7.01	&	6.81	&	6.74	&	 F5V 	\\
8	&	P89-122	&	17301612-2118427	&		&	2012 August 26	&	03:21:33	&		&		&	8.44	&	7.14	&	6.44	&		\\
9	&	P90-022	&	19145561-1810419	&	IRAS 19119-1816	&	2012 September 24	&	03:16:13	&	  	&		&	6.12	&	5.23	&	4.89	&		\\
10	&	P90-081	&	20470928-1223274	&	HD 197928	&	2012 October 23	&	00:19:32	&	10.85	&	9.96	&	8.10	&	7.64	&	7.57	&	 K3V 	\\
11	&	P90-104	&	21412408-0744005	&	LTT 8643	&	2012 October 24	&	02:23:20	&		&	12.20	&	9.39	&	8.75	&	8.57	&	 K5 	\\
12	&	P90-108	&	21423384-0745278	&	HD 206492	&	2012 October 24	&	03:02:12	&	10.00	&	9.12	&	7.43	&	6.97	&	6.84	&	 G5 	\\
13	&	P90-123	&	23184633+0128348	&	StKM 1-2114	&	2012 October 26	&	02:43:17	&		&	12.50	&	9.94	&	9.41	&	9.25	&	 K5 	\\
\enddata
\end{deluxetable}

\section{Results}
\label{section:results}
Table~\ref{tab:results} lists our results, with the
same format as R13. 
The same sequential number used in Table~\ref{tab:list}
is included, along with the 2MASS identification.
The next columns are: the observed rate of the
event; its deviation from the predicted value;
from their combination
the local limb slope $\psi$; and
the observed position angle P.A.
and contact angle C.A., including $\psi$.  We then list
the signal-to-noise ratio (S/N) of the fit to the light curve,
the separation and brightness ratio with their errors, and the two
individual magnitudes based on the total 2MASS magnitude.

\begin{deluxetable}{rlrrrrrrccrr}
\rotate
\tabletypesize{\small}
\tablecaption{Parameters of detected binaries\label{tab:results}}
\tablewidth{0pt}
\tablehead{
\colhead{Seq}&
\colhead{2MASS}&\colhead{$V$(m/ms)}&\colhead{$V/V_{\rm{t}}$--1}&
\colhead{$\psi $($\degr$)}&\colhead{P.A.($\degr$)}&
\colhead{C.A.($\degr$)}&\colhead{S/N}&\colhead{Sep. (mas)}&
\colhead{Br. Ratio}&
\colhead{$K_{\rm 1}$}&\colhead{$K_{\rm 2}$}
}
\startdata
1	&	10304197+0354115	&	0.4823	&	-9.3\%	&	-5	&	252	&	39	&	18.6	&	168.3	$\pm$	1.6	&	2.46	$\pm$	0.01	&	8.86	&	9.84	\\
2	&	17121145-2134332	&	0.7242	&	-1.7\%	&	-5	&	86	&	3	&	212.9	&	5.1	$\pm$	0.2	&	21.8	$\pm$	0.2	&	5.65	&	9.00	\\
3	&	17123952-2136216	&	0.7492	&	3.3\%	&	7	&	287	&	25	&	175.7	&	13.0	$\pm$	0.3	&	1.0729	$\pm$	0.0009	&	6.72	&	6.79	\\
4	&	12460117-0918482	&	0.4382	&	22.1\%	&	7	&	2	&	70	&	62.4	&	8.3	$\pm$	0.4	&	17.7	$\pm$	0.2	&	6.57	&	9.69	\\
5	&	12480144-0908398	&	0.7158	&	-9.2\%	&	-11	&	260	&	-31	&	19.7	&	5.41	$\pm$	0.08	&	2.90	$\pm$	0.05	&	8.08	&	9.24	\\
6	&	17262278-2122494	&	0.5457	&	-0.4\%	&	0	&	117	&	33	&	77.8	&	8.6	$\pm$	0.4	&	39.1	$\pm$	0.9	&	4.49	&	8.47	\\
7	&	17281619-2057493	&	0.4946	&	-4.5\%	&	 3	&	219	&	-47	&	50.9	&	82.6	$\pm$	0.8	&	1.187	$\pm$	0.002	&	7.40	&	7.59	\\
8	&	17301612-2118427	&	0.3095	&	-20.6\%	&	-6	&	134	&	55	&	79.0	&	18.8	$\pm$	0.3	&	22.0	$\pm$	0.1	&	6.49	&	9.84	\\
9	&	19145561-1810419	&	0.6740	&	25.7\%	&	14	&	313	&	64	&	190.5	&	15.2	$\pm$	0.2	&	67.0	$\pm$	0.4	&	4.90	&	9.47	\\
10	&	20470928-1223274	&	0.6285	&	-2.7\%	&	-6	&	66	&	5	&	72.8	&	18.0	$\pm$	0.5	&	10.09	$\pm$	0.05	&	7.67	&	10.18	\\
11	&	21412408-0744005	&	0.6831	&	-1.6\%	&	-4	&	223	&	-15	&	26.7	&	14.3	$\pm$	0.8	&	8.8	$\pm$	0.1	&	8.69	&	11.05	\\
12	&	21423384-0745278	&	0.7241	&	6.6\%	&	10	&	95	&	35	&	77.4	&	4.20	$\pm$	0.03	&	1.54	$\pm$	0.01	&	7.38	&	7.85	\\
13	&	23184633+0128348	&	0.3660	&	-11.4\%	&	-6	&	182	&	-55	&	11.3	&	6.4	$\pm$	2.7	&	1.05	$\pm$	0.02	&	9.97	&	10.03	\\
\enddata
\end{deluxetable}

Due to the selection criteria of our service mode targets
and as already discussed in R13, some of the stars in our
list are in the direction of the Galactic
Bulge. They have generally very 
red colors, which however when plotted in a color-color
diagram
appear consistent with the significant amounts
of interstellar extinction expected in the Bulge.
As a result, a few of our sources have no
optical counterparts and no spectral information.
In the reminder of this section we discuss the individual
cases, following the sequential numbering. 

Star 1:
this is BD+04~2345 = SAO~118343. In spite of the
relatively large projected separation and small brightness ratio
that we detect, it does not appear to have been recorded as a
binary before. It was included  in Tycho-2~\citep{2000A&A...355L..27H} 
and TDSC~\citep{2002A&A...384..180F}.
The TDSC separation detection limit is $0\farcs3$, and we conclude that
the actual separation is also below this value. The alternative
hypothesis, that the
companion is considerably redder (i.e. fainter in the visual) than
the primary, seems less probable in view of the lack of infrared excess
in the colors listed in Table~\ref{tab:list}.

Star 2:
this is an infrared source detected by IRAS (17091-2130) and AKARI.
The source is in the bulge. Its near-IR colors
 appear be be consistent with $\la 5$\,mag of visual
extinction but no visual counterpart is present, possibly pointing
to local (e.g. circumstellar) extinction.
The 2MASS and the DENIS measurements are significantly 
different: $J$=7.083$\pm$0.013 and $K_s$=5.601$\pm$0.022 versus
$J$=6.573$\pm$0.05 and $K_s$=4.705$\pm$0.06, suggesting a variable
nature of the source. Our LO counts are consistent with the 2MASS flux.

Star 3:
this star is SAO~185150 = HD~155469 and 
is listed in multiple visual catalogues, among them HIPPARCOS (HIP~84198),
Tycho-2 and TDSC. However it was not flagged as double in any of them, possibly
because of the close separation between the components. 
Interestingly, a previous
occultation recorded on July 4, 1982 
was reported by \citet{1983AJ.....88.1845E},
also in this case without detection of binarity. This was
a two-channel photoelectric observation in the blue and red, and the
authors quote upper limits on the flux difference of a
possible companion of about 2.8\,mag (our value is $< 0.1$\,mag).
Their observation and ours were recorded along
P.A. differing by 58$\degr$, and are separated by
almost exactly 30 years: this is significantly longer
than the period that we estimate below.
SAO~185150 was included in a catalog of empirical angular diameters
by ~\citet{2006A&A...450..735M} who
estimate a value of $0.252$\,mas.
We can only put an 
upper limit of 2\,mas on the angular size of the components,
\citep[see][]{apjssur}, but we note that if they had significantly
different colors a discrepancy would have been noticed in the
spectral energy distribution, especially in consideration of the
high accuracy claimed by the authors (1\% in effective temperature).

\begin{figure}
\includegraphics[angle=-90.0, width=7.5cm]{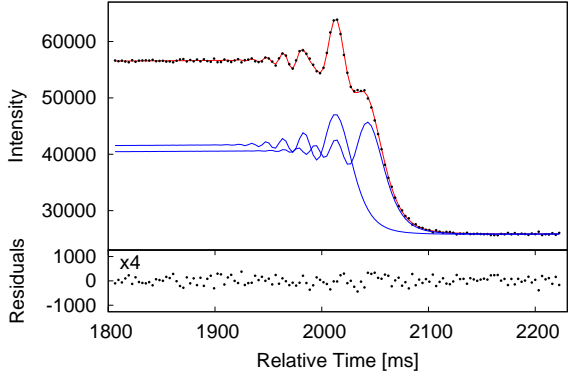}
\caption{Occultation light curve (dots) and
best fit (solid line) for HD~155469, showing the 
first-time detection as a binary star. For illustration,
the light curves of the two components are also shown.
The $\chi^2_n$ of the binary fit is 0.97.
The lower panel shows the fit residuals,
enlarged for clarity.
}
\label{fig:hd155469}
\end{figure}

Contrary to the above observations and estimates, 
Fig.~\ref{fig:hd155469} clearly shows the binary
nature of SAO~185150. The lack of detection in the Tycho and previous LO
data suggests that projection effects should not be very large. For the sake
of discussion, we assume the average de-projection factor of $\pi / 2$, i.e.
a true separation of $\approx 20$\,mas.
SAO~158150 is classified as
a F5 dwarf, and consistently with its apparent brightness
Hipparcos measured a parallax corresponding to 86\,pc \citet{2007A&A...474..653V}
which would imply an actual separation of $\approx 2$\,AU.
We find a flux ratio close to unity, and assuming a total mass of $\approx 2$\,M$_\odot$
and a circular orbit the period should be of the order of very few years.
SAO~185150 appear to be an ideal candidate for an orbital solution when
observed by adaptive optics in short wavelengths, with the potential to lead
to accurate dynamical masses.

Star 4:
this is BD-08~3423 = SAO~138958, a G4 dwarf. It is
listed in HIPPARCOS, Tycho-2 and TDSC without
mention of binarity.
It is a high-proper motion star (LTT~4882), with $\mu$(RA) = $311.6\pm0.7$
and 
$\mu$(DEC) = $-219.2\pm 0.5$ mas\,yr$^{-1}$, and with a parallax 
that places it at about 50\,pc
\citep{2007A&A...474..653V}. It was also included in the work
by \citet{2006A&A...450..735M} mentioned for star 3.
This system appears a good candidate for a rapid
orbital determination: a rough estimate based on $\approx 1$~M$_\odot$
and a (deprojected) $\approx 0.7$~AU separation points to a period
of less than one year.

Star 5:
this is HD~111241 = SAO~138971. Similarly to the previous stars,
it is listed in Tycho-2 and TDSC but without binarity flags, 
most likely due to the very close separation between the components.
We note that the counts from our LO curve are consistent with
a flux almost double than the 2MASS $K_s$ magnitude.

\begin{figure}
\includegraphics[angle=-90.0, width=7.5cm]{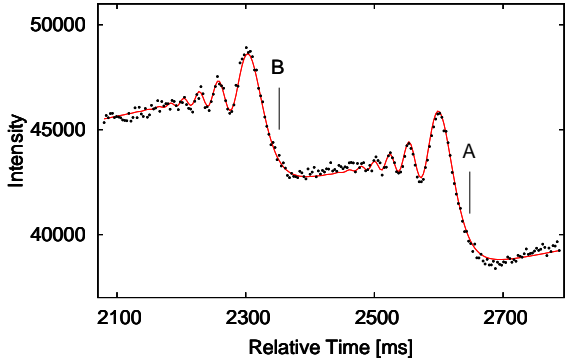}
\caption{Occultation light curve (dots) and
best fit (solid line) for HD~158122, the only previously
known binary star in our list.
The times of
geometrical occultation of the two components
are marked. 
}
\label{fig:hd158122}
\end{figure}

Star 6:
this
is an infrared source measured by
AKARI 
and IRAS.
It is the brightest source in our list with $K_s=4.5$\,mag,
and in fact we observed it with a narrow band filter to avoid
possible non-linearities. It shows very red colors, which are
however consistent with its location in the Galactic Bulge.

Star 7:
this is HD~158122, 
 a well known visual binary \citep[ADS 10561 AB,][]{aitken1932}.
Its orbit was derived for the first time by \citet{baize1950}  which
gave a period of  36.7 years, and a spectral type of the primary of F8.
This system has
been extensively observed
by means of speckle interferometry \citep[see][and references therein]{hart2000}.
Additional measurements were published by \citet{mason2009}
and
\citet{toko2010}.
The Sixth Catalog of Orbits of Visual Binary Stars \citep{hart2009}
reports an orbital solution which is graded as good, 
with a period of 45~years. Using these elements
\citet{cvet2010} determined the dynamical masses 
as $1.24$ M$_\odot$ and $1.21$ M$_\odot$  
for the primary and secondary, 
respectively, with  spectral type F6 for both stars.

Our light curve and best fit are shown in Fig.~\ref{fig:hd158122}.
Using the same orbital elements mentioned above, we estimate that
at the date of the occultation the P.A. was $196\degr$ and the
separation $0\farcs105$. Our measured projected separation 
appears smaller than the expected one (83 instead of 97\,mas), but
approximately in agreement given possible uncertainties in
the orbital elements and also in our predicted P.A.
Our LO-derived brightness ratio in the K band
is an additional first-time constraint to the luminosity of the
two components. 

Star 8: 17301612-2118427 does not have a known
optical counterpart. It is the second reddest object
in our list with $J-K=2.0$\,mag. We note that our
counts show only about half of the flux measured
by 2MASS, pointing to variability.

Star 9:
19145561-1810419 is an infrared source detected by IRAS, AKARI and 
WISE. It was listed in Tycho-2 with no report for binarity, and the 
measured proper motions are statistically undistinguishable from 
zero. One 2MASS and two DENIS measurements are 
available in J and $K_s$, 
 but they differ significantly (in the range 5.4 to 6.1\,mag,
and 4.0 to 4.9\,mag in the two bands respectively, with small errors),
suggesting a variable 
nature of the source. Also \citet{2005AcA....55..275P} 
report 0.15\,mag $V$-band amplitude. 
\citet{2006ApJ...638.1004A} determined an 
effective temperature T$_{\rm eff}=5958^{+751}_{-772}$; however,
\citet{2008PASP..120.1128O} estimated a cooler spectral class of M5\,III; 
\citet{2010PASP..122.1437P}  also derived a spectral type M5\,III and 
a distance of $\sim$1.6\,kpc. The variability, together with the 
mid-IR detections indicate that this may be an AGB star. Finally, 
the object was detected by GALEX.

\begin{figure}
\includegraphics[angle=-90.0, width=7.5cm]{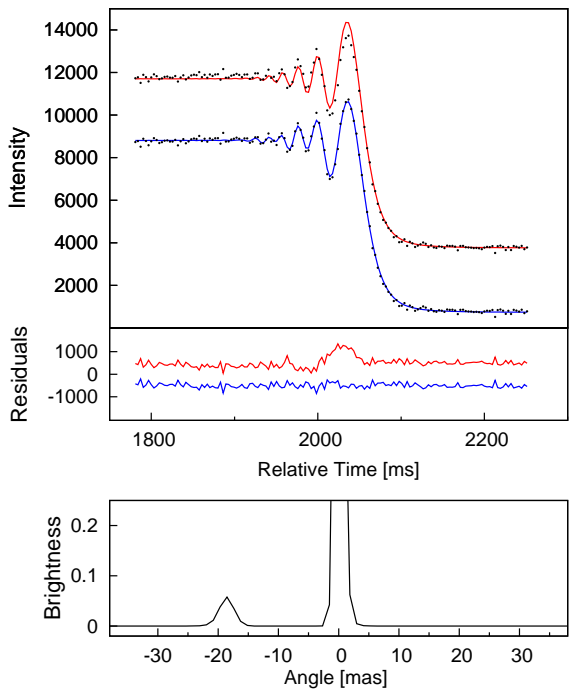}
\caption{Upper panel: occultation data (dots) for
HD~197928, a newly detected binary, with best fits
by a single (above) and a binary star (below).
The middle panel shows the residuals for the two cases,
shifted by $\pm 500$ counts for clarity. 
The $\chi^2_n$ values for the single and the binary source
fits are 3.42 and 1.06, respectively. 
The model-independent brightness profile is shown
in the bottom panel. The profile of the central star
is truncated in this enlarged view.
}
\label{fig:hd197928}
\end{figure}

Star 10:
this is HD~197928, a nearby high-proper motion dwarf star, listed 
in FOCAS-S, HIPPARCOS (HIP~102569), and Tycho-2 but with no mention of possible
binarity.  
The star
was detected also by WISE and GALEX.
Objective prism 
spectra yielded a spectral type K2V \citep{1972AJ.....77..486U}. 
\citet{2006ApJ...638.1004A} determined an 
effective temperature T$_{\rm eff}=4938^{+74}_{-122}$, metallicity 
[Fe/H]=$-$0.12$^{+0.32}_{-0.24}$, and a distance D=57$^{+53}_{-21}$\,pc; 
\citet{2008PASP..120.1128O} estimated a spectral class K0\,III but the
luminosity class seems inconsistent with
the measured parallax; 
\citet{2010PASP..122.1437P} derived spectral type K3\,V and a distance 
D$\sim$42\,pc; \citet{2011MNRAS.411..435B}, estimated somewhat higher 
effective temperatures T$_{\rm eff}=5879^{+486}_{-505}$ or 
T$_{\rm eff}=5604^{+293}_{-357}$, depending on the adopted models. 
Interestingly, \citet{2006ApJ...638.1004A} give nearly zero 
probability for binarity, based on the quality of their SED fit. 
A model-independent fit to the data \citep[CAL method,][]{CAL}
already revealed the presence of a companion, as shown
in the bottom panel of 
Fig.~\ref{fig:hd197928}. The model-dependent fit confirmed
this and in fact rules out convincingly the single star hypothesis.

Star 11:
this is a nearby high-proper motion star, first identified
by \citet[LTT 8643]{1957QB6.L98........}. In the revised NLTT 
\citet{2003ApJ...582.1011S} report $\mu$(RA) = $143.4\pm5.5$ and
$\mu$(DEC) = $203.8\pm5.5$ mas\,yr$^{-1}$. Kuiper classified it as K4/5\,V 
(Bidelman 1985), and \citet{1986AJ.....92..139S} as K5\,V. The 2MASS 
and DENIS magnitudes agree, within the errors. \citet{2010PASP..122.1437P} 
derived a spectral type K7\,V and a distance of $\sim$64\,pc. The object 
was detected by WISE, GALEX, and SDSS.

\begin{figure}
\includegraphics[angle=-90.0, width=7.5cm]{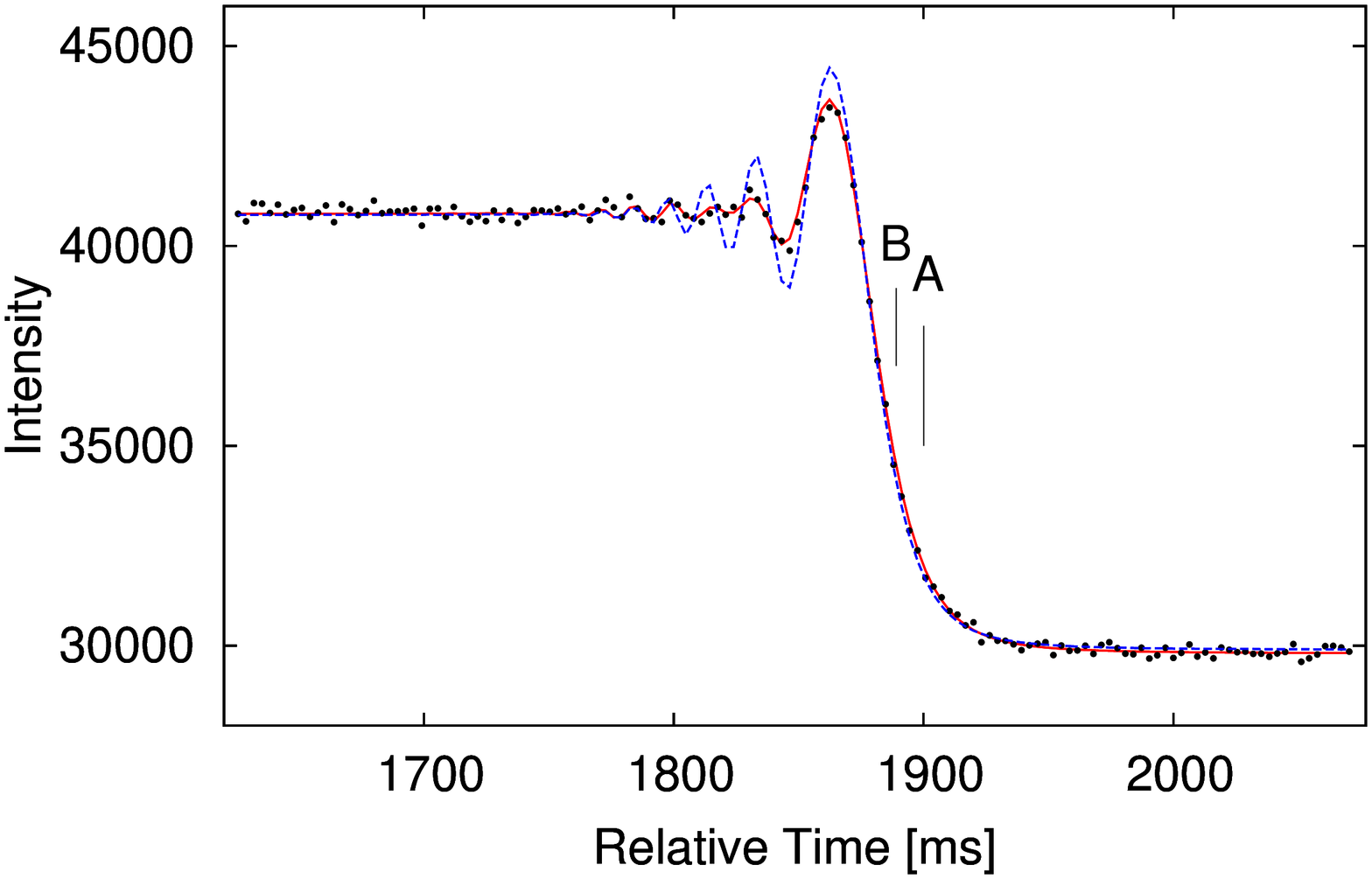}
\caption{Occultation data (dots) for
HD~206492, a newly detected binary with the smallest 
projected separation among our results. The best fit with
a binary source model is shown a solid line, with segments
of length proportional to the intensity of each
component marking their times of
geometrical occultation.
For comparison, also the best fit by an unresolved single
source is shown as a dashed line.
The $\chi^2_n$ values for the single and the binary source
fits are 2.90 and 0.75, respectively.
}
\label{fig:hd206492}
\end{figure}

Star 12:
HD~206492 = 	SAO~145602
 was listed in both FOCAST-S \citet{1994BICDS..44....3B} 
and Tycho-2 catalogues as an object with small, but measurable proper 
motion: $\mu$(RA) = $11.2\pm1.7$ and $\mu$(DEC) = $2.2\pm1.9$ mas\,yr$^{-1}$ 
\citep{1998A&A...335L..65H}.
There is no mention of possible binarity. Objective 
prism spectra yield a spectral type G8\,III \citep{1999mctd.book.....H}. 
\citet{2006ApJ...638.1004A} 
determined an effective temperature T$_{\rm eff}=4867^{+223}_{-125}$, 
and a distance D=177$^{+195}_{-93}$\,pc; \citet{2008PASP..120.1128O} 
estimated a spectral class G5\,III; \citet{2010PASP..122.1437P} 
derived a spectral type K2\,V and a distance D$\sim$38\,pc. The object 
was detected by WISE, GALEX, and SDSS. \citet{2007KFNT...23..102R} 
suggested that the object may be a local Red Clump star, base on its 
$K_s$-band reduced proper motion.
Our light curve and best fit are shown in Fig.~\ref{fig:hd206492}.
Our counts show about 30\% flux decrease with respect to 2MASS.

Star 13:
this 
is a nearby K5 dwarf
with small 
proper motion \citep{1986AJ.....91..144S}.
The 2MASS and DENIS colors agreee, 
within the errors, and also our LO counts agree with the 2MASS
magnitude: the source does not seem to exhibit large variability. 
The object was detected by WISE, GALEX, SDSS, and 
in some UKIDSS filters.

\section{Conclusions}
\label{sec:conclusions}
With the observations reported here, we have
 exhausted the sample of LO observed
at the VLT with ISAAC, which is about to be decommissioned.
A total of 1226 events were recorded over almost exactly
7 years (with gaps of up to a year in between). 
Discarding the negative detections and the grazing events,
we could analyze 1161 light curves. The majority of these
were found to be unresolved \citep{apjssur}, and several
showed an extended appearance. Many in this latter class
are still awaiting publication. Concerning binary and
multiple systems, with the present work we have
totalled 97 stars, or a detection rate of
$\approx 8$\% among randomly selected stars. Of these,
90 were first-time detections.

Specifically for this paper, 
from the first to the last
night considered in Table~\ref{tab:list},
a total of 290 LO light curves were observed, 
therefore our binary
detection fraction is (13/290) $\approx 4.5$\%, which
is less than value above and also less than 
the 6.5\% rate reported in R13. As already
discussed  in this latter work, the sky distribution of
the sources may affect indirectly their
average distance. For an equal distribution of separations,
more distant stars will be harder to be detected
as binaries.

As was the case
in R13, half of
the binary stars in this paper have 
flux ratios in the range 2$\le \Delta{\rm K} \le$4.6\,mag.
This makes them challenging for long-baseline interferometry.
As for the separations, even allowing for a deprojection factor,
they are largely unaccessible for current very large telescopes
even with adaptive optics.
The follow-up of most of these systems will thus be
possible mainly with future extremely large telescopes.
Concerning astrometric confirmation and follow-up, the majority
of our newly detected systems was beyond the capabilities
of Hipparcos, but they 
should constitute a
valuable benchmark for the validation and performance
assessment of more accurate future observations, such as made
possible by GAIA. 

\acknowledgments
We are grateful to the ESO staff in Europe and Chile for
carrying out the service mode observations.
OF acknowledges financial support from MINECO through
a {\it Juan de la Cierva} fellowship and 
from \emph{MCYT-SEPCYT Plan Nacional I+D+I AYA\#2008-01225}.
This research made use of the Simbad database,
operated at the CDS, Strasbourg, France, and
of data products from the Two Micron All Sky Survey (2MASS), 
which is a joint project of the University of Massachusetts 
and the Infrared Processing and Analysis Center/California Institute 
of Technology, funded by the National Aeronautics and 
Space Administration and the National Science Foundation.
This research has made use of the Washington Double Star Catalog
maintained at the U.S. Naval Observatory.

\clearpage

\end{document}